# Supporting Tool for The Transition of Existing Small and Medium Enterprises Towards Industry 4.0

Miguel Baritto*
Department of Mechanical Engineering
University of Alberta, Edmonton
116 St & 85 Ave, Edmonton, AB
T6G 2R3, Canada
baritto@ualberta.ca

Md Mashum Billal*
Dept. of Mechanical Engineering
(Engineering Management)
University of Alberta, Edmonton
116 St & 85 Ave, Edmonton, AB
T6G 2R3, Canada
mdmashum@ualberta.ca

S. M. Muntasir Nasim
Department of Mechanical Engineering
University of Alberta, Edmonton
116 St & 85 Ave, Edmonton, AB
T6G 2R3, Canada
snasim@ualberta.ca

Rumana Afroz Sultana
Department of Mechanical Engineering
University of Alberta, Edmonton
116 St & 85 Ave, Edmonton, AB
T6G 2R3, Canada
rumanaaf@ualberta.ca

Mohammad Arani
Dept. of Systems Engineering
University of Arkansas at Little Rock
2801 S. University Ave., Little Rock, AR 72204, USA
mxarani@ualr.edu

Ahmed Jawad Qureshi
Department of Mechanical Engineering
University of Alberta, Edmonton
116 St & 85 Ave, Edmonton, AB
T6G 2R3, Canada
ajqureshi@ualberta.ca

*Abstract*—The rapid growth of Industry 4.0 technologies such as big data, cloud computing, smart sensors, machine learning (ML), radio-frequency identification (RFID), robotics, 3D-printing, and Internet of Things (IoT) offers Small and Medium Enterprises (SMEs) the chance to improve productivity and efficiency, reduce cost and provide better customer experience, among other benefits. The main purpose of this work is to propose a methodology to support SMEs managers in better understanding the specific requirements for the implementation of Industry 4.0 solutions and the derived benefits within their firms. A proposed methodology will be helpful for SMEs manager to take a decision regarding when and how to migrate to Industry 4.0.

## I. INTRODUCTION

Industry 4.0, also called the fourth Industrial Revolution, is a flourishing research topic currently. It is the combination of several emerging tools and technologies, for example, 3D-printing, big data, cloud computing, smart sensors, machine learning (ML), radio-frequency identification (RFID), robotics, and Internet of Things (IoT), among others [1]. These advanced tools can integrate people, machines, and data for creating more agile and responsive supply chains, improving productivity.

Implementing Industry 4.0 solutions has become familiar today in many process manufacturing organizations, as it has differentiated of tools and technologies to solve critical problems. However, the novelty and diversity of tools making up this industrial revolution represent an obstacle for the transition towards Industry 4.0, above all for Small and Medium Enterprises (SMEs), which are not willing to take the risk of investing in a journey full of uncertainties [2], and usually do not have the initial capability, in terms of software, hardware, skilled personnel, and organization, to undertake the transition.

This study aims to provide guidelines to help SMEs managers to decide transition towards industry 4.0, by identification of specific benefits of conceptualized Industry 4.0 projects tailored to the current reality of the company.

## II. LITERATURE REVIEW

The literature review is divided into three parts. First, it is reviewed how SMEs are approaching to Industry 4.0, to identify a research gap. Second, and to the identified research gap, different existing tools for Industry 4.0 assessment are examined, and its suitability for application to SMEs is discussed. Finally, some works regarding specific applications of Industry 4.0 are presented, as examples of the portfolio of projects that could be used in the proposed methodology, as detailed in the next section.

### A. Reality of SMEs facing Industry 4.0

There is a significant relationship between company size and the implementation of Industry 4.0. In general, large enterprises are more advanced in integrating IT systems into their operations than medium-sized enterprises, and the latter are more advanced than small enterprises [3]. This trend corresponds to the specific reality of SME in opposition to Multinational Enterprises (MNEs), as explained by Mittal et al. [4] and summarized in TABLE I. . Schröder [3] also identified obstacles for Industry 4.0 implementation in SMEs and mapped them back to the nature of these kinds of enterprises, for example, lack of digital strategy, resource scarcity, lack of standards, and poor data security. Kleindienst and Ramsauer [5] commented on the problems SMEs are facing in the Industry 4.0 context. These authors identified the lack of specialized IT personnel as the biggest of those problems since usually SMEs develop very easy spreadsheet-based tools (tailored solutions) with limited connectivity with other resources within the company or by interacting with other companies, making too

---

difficult the implementation of Industry 4.0. Another big obstacle identified by [5] is the SMEs lack of skills to identify where to gather data across the entire supply chain and value chain, and how to run and maintain such system of sensors.

Müller et al. [6] explored how SMEs approach implementation of Industry 4.0 and how they could innovate in their business to increment the derived benefits. Managers from 68 German SMEs were interviewed regarding the understanding of the term Industry 4.0, challenges for implementation, and Industry 4.0-enabled innovations across the business model. Although the majority of the managers were familiar with the concept, they evidenced difficulties to understand how to create value from the data collection, stressing out that the big challenge is the high investment for machine digitalization, standardization, and networking between suppliers and customers, with little benefit in return. Schuh et al. [7] and Erol et al. [2] also commented on the struggling of companies in bringing Industry 4.0 ideas down to the shop floor, mainly because the derived increment on productivity is hard to identify. From the cited studies it is evident the SMEs need for a guideline that supports managers to offset the identified obstacles by identifying benefits of implementing Industry 4.0 within their firms. Available supporting tools where reviewed in the next part of the literature review.

TABLE I. COMPARISON OF SMEs AND MNEs BASED ON FEATURES [4]

| Feature | SMEs | MNEs |
| --- | --- | --- |
| Financial Resources | Low | High |
| Technical Resources | Low | High |
| Software Umbrella | Tailored solutions | Standardized solutions |
| R&D | Low | High |
| Standards compliance | Low | High |
| Organizational culture | Less complex | Complex |
| Employee engagement | Specific domain | Multiple domains |

### B. Existing Industry 4.0 readiness tools and their implementation in SMEs

According to Pöppelbuß and Röglinger [8] (cited in Gajsek et al. [9]), a maturity model includes a sequence of levels that form a logical path from an initial state to maturity. The definition of the levels depends on the authors. For example, Qin et al. [10] define five Industry 4.0 maturity levels: single-station automated cells, automated assembly systems, flexible manufacturing systems, computer-integrated manufacturing, and reconfigurable manufacturing system. Mittal et al. [4] performed a critical review of Industry 4.0 maturity models and the implication for SMEs. They concluded that just a few of the reviewed maturity models take into consideration the specific requirements and challenges of SMEs. The "Level 1" of those models usually implies that the enterprise already has the resources and skills required to move to the next level in the transition towards Industry 4.0. This kind of "Level 1" is more suitable for MNEs but tends to be beyond the current reality of SMEs, and for this reason, they suggested the implementation of a "Level 0" to address basic requirements, for example, the recognition of technologies that were not relevant for the business, so far.

An example of maturity models with a "Level 1" too high for SMEs is found in Schumacher et al. [11]. They combined the evaluation of technological capacities and organizational strategies into nine dimensions (Products, Customers, Operations, Technology, Strategy, Leadership, Governance, Culture, and People) for assessing Industry 4.0 readiness of manufacturing enterprises. A maturity index for each dimension is defined based on the response to a 62 items survey. One of the questions used in the questionnaire is: do you use a road map for the planning of Industry 4.0 activities in your enterprise? This means that the enterprise has already made some steps towards Industry 4.0, which is not necessarily the case for SMEs. Although the results of this kind of assessment provide to manager levels of information about what dimensions should be improved to reach Industry 4.0, is not useful to identify areas within the enterprise that would be beneficiated by the transition.

Another example of such tools with a high entry level is found in Ghobakhloo et al. [12] who developed a strategic roadmap for manufacturers for the transition to Industry 4.0. This roadmap guides the transition process in six organizational areas: Strategic Management, Marketing Strategy, Human Resources Strategy, IT Maturity Strategy, Smart Manufacturing Strategy, and Smart Supply Chain Management Strategy. The roadmap supports the company strategy definition to face the transition process, but again, it supposes that the decision to undertake such transition has been already taken, and does not provide guidelines on how to identify benefits derived from Industry 4.0 as a starting point.

Regarding tools that consider SMEs specific requirements, Kleindienst and Ramsauer [5] proposed a three-step procedure model to identify the Industry 4.0 need-for-action for SMEs. The first step consists in data gathering for calculation of KPIs in the second step. The KPIs should be chosen in such a way they can be calculated using data available in SMEs. The third step corresponds to deciding for priorities in terms of five key Industry 4.0 topics: horizontal integration over value-creation networks, consistency of engineering over the whole lifecycle, vertical integration and cross-linked production systems, new social infrastructures of work, and continuous development of cross-sectional technologies. Although the KPIs approach offers an SME tailored tool, these authors do not offer details about which KPIs are suitable for this kind of study.

Ganzarin and Errasti [13] proposed a methodology for Industry 4.0 collaborative diversification, based on the strategic guidance towards Industry 4.0 presented by [2]: Vision stage, dedicated to defining a tailored Industry 4.0 vision to develop an own general understanding of the concept within company-specific capabilities and resources; Roadmap stage, in which the company defines a technology portfolio and capabilities needed to be based on the vision defined in the previous step; and Projects, which include training capacitation and risk management of specific Industry 4.0 projects .

Mittal et al. [14] developed a four-step Smart Manufacturing (SM) adoption framework for SMEs: identify manufacturing data present in the SME; SM readiness assessment of data-hierarchy; develop a smart manufacturing tailored vision for SME; Identify tools and practices to realize tailored SM vision. Although this framework is strongly SMEs focused, the approach still relies on existing data collection systems and previous awareness of managers on Industry 4.0 solutions, therefore it is not clear how to apply the framework to companies that do not count with initial capabilities for the transition towards Industry 4.0. Also, according to the case studies presented in this study, it seems that this framework provides guidelines for implementation of Industry 4.0 solutions already envisaged by SMEs managers, rather than help them to generate the solutions.

*C. Some applications of Industry 4.0*

The KPI that SMEs expect to improve after investing in new technology (e.g. Industry 4.0 solutions) include cost, quality, flexibility, productivity, and delivery lead time ([15]–[17]). Moeuf et al. [17], literature review of existing applied research covering different Industry 4.0 issues concerning SMEs and found that flexibility and productivity are the most frequent KPI targeted by Industry 4.0 in SMEs and that this type of enterprise does not exploit all the resources for implementing Industry 4.0, usually limiting the solutions to the adoption of Cloud Computing and the Internet of Things. A summary of their finds is shown in TABLE II. .

TABLE II. INDUSTRIAL PERFORMANCE OBJECTIVES TARGETED BY INDUSTRY 4.0 ON SMEs, BASED ON [17]

| KPI improvement | Industry 4.0 tools |
|---|---|
| Flexibility | • Cloud computing |
|  | • Production planning optimization |
| Productivity | • Production flow based on IoT |
| Delivery lead time | • Cloud computing |
|  | • IoT-based Lean Manufacturing |
| Cost reduction | • 3D digital models |
|  | • Real-time management of flows |
| Quality | • RFID |

TABLE III. shows some examples of Industry 4.0 specific applications in SMEs. This kind of portfolio of projects could help SMEs managers to visualize how Industry 4.0 brings concrete benefits to their production processes. For example, real data acquisition coupled with rescheduling algorithms serves to adapt the schedule to unplanned events, ensuring the continuity of operations [18]; augmented reality could be used for enhanced Go &-See by displaying right in place the status of operations [19]. In this way, process bottleneck can be readily identified; strategies for reducing conversion costs, like scrap cost and consumables cost, can be generated by applying quality control tools based on real-time data acquired via Cyber-Physical System (CPS) [20]; smart sensors can be used to detect and automatically resolve safety issues, like leaking of dangerous substances, or even for automatic order placement [21]; robotics intelligent self-driving vehicles have been used for material handling in manufacturing plant and warehouses, reducing the delivery time of final products to the customers [22]; RFID is a wireless communication technology that is used to automatic identification, tracking and data collection from any tagged object in a supply chain operating environment by transmitting radio signals [23], and also can be combined with IoT to get production feedback in real time [24].

TABLE III. APPLICATION OF INDUSTRY 4.0 IN SMEs

| Application | Tools | References |
|---|---|---|
| Rescheduling | Real data processing/cloud computing | [18] |
| Debottlenecking | Augmented reality | [19] |
| Manufacturing cost reduction | CPS/cloud computing | [20] |
| Safety | Smart sensor/IoT | [21] |
| Order placement | Smart sensor/IoT | [21] |
| Decreasing delivery time | Intelligent self-driving vehicles | [22] |
| Tracking items in the Supply Chain | RFID | [23] |

### III. PROBLEM STATEMENT

One of the reasons that delay the implementation of Industry 4.0 in SMEs is the challenge to understand the benefits in terms of productivity, competitiveness, and value creation that Industry 4.0 would bring to the firms. SMEs managers think that moving into Industry 4.0 represents a high-cost investment that customers are not willing to pay for it since at least at the beginning, it generates only information, not solutions [6]. These beliefs are driven by the reality of SMEs (e.g. different automation degrees of pieces of machinery, if any; lack of capability for total digitalization; budget constraints; lack of skilled personnel) that makes the transition towards Industry 4.0 looks like a titanic endeavor. For an extended implementation of Industry 4.0 in SMEs, managers should count on guidelines for identification and assessment of value-creation opportunities tailored to the reality of their firm, supporting them in deciding to start the transition. However, the existing frameworks, roadmaps, and maturity models are more suitable for large MNEs that have already made the decision to move towards Industry 4.0 and even have dedicated resources and infrastructure (e.g. connected machines, sensors, IT department, etc.) as a starting point in the journey [4], [6]. The reality of SMEs makes impractical the application of those tools, being this the issues addressed in this work.

In that sense, the objective of this research project is to propose a framework to help SMEs managers identifying specific opportunities and concrete projects that depict the benefits of the transition and ease the engagement with Industry 4.0. By applying the proposed framework, it is expected to visualize the benefits that the company would derive from Industry 4.0, such as adding value to the supply chain, minimization of cost, broaden business control, and increment of profitability.

### IV. PROPOSED METHODOLOGY

Fig. 1 illustrates the proposed methodology for generating Industry 4.0-based solutions according to the current state of the company. The goal is to generate a set of proposals, or pilot projects, for the implementation of Industry 4.0 tools that respond to current needs and improvement opportunities previously identified. By doing this, SMEs managers would be able to apply existing maturity models to identify the areas of the company that should be improved to meet the requirements of specific projects. The proposed methodology is comprised of four gates and three processes to advance through the gates.

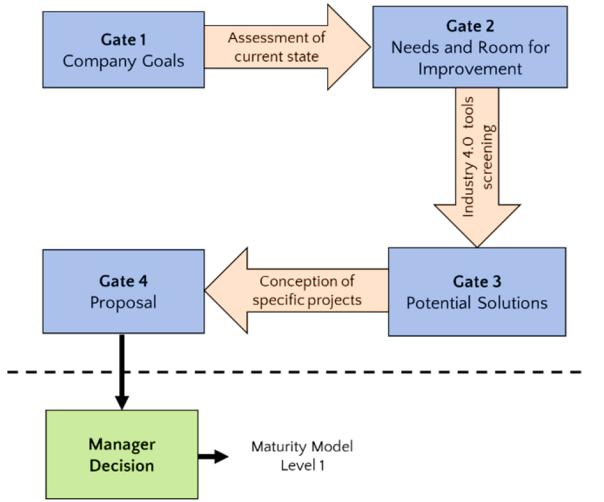

Fig. 1. Schematic of the proposed methodology

*A. Gate 1: Company Goals*

The starting point of the methodology is the identification and understanding of company goals, which represent the target state of the company strategy. Industry 4.0 solutions must be aligned to such a strategy to effectively add value to the business. This gate aims to keep SMEs focussed on the company goals during the generation of Industry 4.0 solutions, avoiding stray ideas and forcing the adoption of Industry 4.0 just because of the hype of this new concept. Moreover, in case the methodology is applied by an external consultant, this gate ensures the overall outcome meets company goals, generating tailored proposals with a high chance of implementation, rather than generic solutions that would be discarded because of their low applicability in the company context.

*B. Form Gate 1 to Gate 2: Assessment of Current State*

During the process, current data is compiled and analyzed to determine the current state of the company. Based on [15]–[17], it is recommended to assess the current flexibility, productivity, costs, quality, and delivery lead time. Other indicators that could be considered include sales, selling price, inventory, and market share.

*C. Gate 2: Needs and Room for Improvement*

The second gate is the needs and room for improvement, which are identified as the deviation of the current state from the company goals. Depending on the managerial culture of the company, SMEs managers could be already aware of the needs and areas that can be improved, thus this gate would be the starting point of the methodology, rather than gate 1.

*D. Form Gate 2 to Gate 3: Industry 4.0 Tools Screening*

This process implies the identification of Industry 4.0 tools that can be applied to address the needs and room for improvement previously identified. The search for Industry 4.0 tools is driven by the specific context of the company, that is, SMEs managers start from a particular need within the company and look for tools to satisfy that need, rather than start with a tool and look how to fit it within the company.

Because the uncertainties around the applicability of Industry 4.0 could bias the adoption of tools, it is recommended to perform a screening of technologies already implemented or with a Technology Readiness Level (TRL) as high as possible, in order to build up confidence towards the project that would result after applying this methodology. Industrial and academic journals and published case studies are good sources of such Industry 4.0 tools.

*E. Gate 3: Potential Solutions*

The third gate is the potentials solutions, identified by matching each of the needs and opportunities with Industry 4.0 tools found during the screening process. Often, a need could be solved by implementing one of several suitable Industry 4.0 solutions, or a combination of diverse tools, for this, the aim of this gate is to discard those infeasible solutions in terms of applicability criteria, for example, the requirement of energy consumption is prohibitive in remote areas. Critical thinking, expert judgment, and brainstorming are some of the resources that could be employed during this creative process.

*F. From Gate 3 to Gate 4: Competition of Specific Projects*

Once the potential solutions have been identified, it is possible to conceptualize specific projects for each solution. The idea is to generate all the pieces of information that could help the SMEs managers to understand why and how to implement the potential solutions through the execution of specific projects. Some pieces of information are flow process diagrams, list of requirements (e.g. software, hardware, organizational, skilled workers), and feasibility estimation such as order the magnitude Techno-economic Assessment and risk analysis.

*G. Proposal*

The final gate corresponds to a document compiling an executive summary and all the information generated in the previous process. This document can be used by SMEs managers in making decisions regarding which projects should be considered for execution, returned for improvement or clarification, or shelved. For those projects considered for execution, SMEs manager can now apply some of the existing maturity models to evaluate the actions required for the implementation of the specific solutions

## V. Conclusion

The objective of this study was to provide a guideline to help SMEs managers to make a decision in the transition towards industry 4.0, by identification of specific benefits of conceptualized Industry 4.0 projects tailored to the current reality of the company. The direction of this research project was to propose a framework to help SMEs managers in identifying specific opportunities and concrete projects that depict the benefits of the transition and ease the engagement with Industry 4.0.